\documentclass[twoside]{article}

\usepackage[accepted]{aistats2019}

\usepackage{microtype}
\usepackage{graphicx}
\usepackage{subfigure}
\usepackage{booktabs} 

\usepackage{algorithm}
\usepackage{algpseudocode}

\usepackage{natbib}
\setcitestyle{authoryear,round,citesep={;},aysep={,},yysep={;}}

\usepackage{color}
\definecolor{mydarkblue}{rgb}{0,0.08,0.45}
\usepackage[colorlinks=true,allcolors=mydarkblue]{hyperref}

\usepackage{appendix}
\usepackage{amsmath}
\usepackage{xspace}
\usepackage{multicol}
\usepackage{afterpage}
\usepackage{capt-of}
\newcommand{\ee}{\text{rest}}
\newcommand{\boldx}{\mathbf{x}}
\newcommand{\boldy}{\mathbf{y}}
\newcommand{\boldz}{\mathbf{z}}
\newcommand{\boldu}{\mathbf{u}}
\newcommand{\boldU}{\mathbf{U}}
\newcommand{\boldV}{\mathbf{V}}
\newcommand{\boldv}{\mathbf{v}}
\newcommand{\boldX}{\mathbf{X}}
\newcommand{\boldZ}{\mathbf{Z}}
\newcommand{\boldw}{\mathbf{w}}

\newcommand{\N}{\mathcal{N}}
\newcommand{\dist}{\text{d}}
\newcommand{\crossovers}{ensemble crossover scheme\xspace}

\begin{document}

\runningtitle{Augmented Ensemble MCMC sampling in Factorial Hidden Markov Models}

\twocolumn[

\aistatstitle{Augmented Ensemble MCMC sampling \\ in Factorial Hidden Markov Models}

\aistatsauthor{Kaspar M\"artens \And Michalis K. Titsias \And  Christopher Yau}

\aistatsaddress{University of Oxford\\kaspar.martens@stats.ox.ac.uk \And Athens University of\\ Economics and Business\\mtitsias@aueb.gr \And Alan Turing Institute\\University of Birmingham\\c.yau@bham.ac.uk} 

]

\begin{abstract}
Bayesian inference for factorial hidden Markov models is challenging due to the exponentially sized latent variable space. Standard Monte Carlo samplers can have difficulties effectively exploring the posterior landscape and are often restricted to exploration around localised regions that depend on initialisation. We introduce a general purpose ensemble Markov Chain Monte Carlo (MCMC) technique to improve on existing poorly mixing samplers. This is achieved by combining parallel tempering and an auxiliary variable scheme to exchange information between the chains in an efficient way. The latter exploits a genetic algorithm within an augmented Gibbs sampler. We compare our technique with various existing samplers in a simulation study as well as in a cancer genomics application, demonstrating the improvements obtained by our augmented ensemble approach.  
\end{abstract}

\section{Introduction}
 
Hidden Markov models (HMMs) are widely and successfully used for modeling sequential data across a range of areas, including signal processing \citep{crouse1998wavelet}, genetics and computational biology \citep{marchini2010genotype, yau2013oncosnp}. The HMM assumes that there is an underlying unobserved Markov chain with a finite number of states, which generates a sequence of observations $y_{1:T} := (y_1, \ldots, y_T)$ via a parametric emission distribution. Inference over the latent sequence $x_{1:T}$ and the parameters can be carried out either from a likelihood \citep{rabiner1986introduction} or Bayesian \citep{scott2002bayesian} perspective. In the latter, conditional sampling can be used where the parameters and latent sequences are updated iteratively conditional on the other being fixed. Latent sequences can be sampled using forward-filtering-backward-sampling (FF-BS) \citep{scott2002bayesian}. 

The Factorial HMM (FHMM) \citep{ghahramani1997factorial} is an extended version of the standard HMM where instead of a single latent chain, there are $K$ latent chains. That is, given observations $y_{1:T}$, our goal is to infer a $K \times T$ latent matrix $\boldX = (\boldx_1, \ldots, \boldx_T)$ whose columns evolve according to Markov transitions. Here we focus on the case where $\boldX$ is binary, in which case the element $x_{k, t}$ indicates whether latent feature $k$ contributes to observation $y_t$. 
The joint distribution $p(y_{1:T}, \boldX)$ is given by
\begin{align*}
p(y_{1:T}, \boldX) &= \left( \prod_{t=1}^T p(y_t | \boldx_t) \right) \left( p(\boldx_1) \prod_{t=2}^T p(\boldx_t | \boldx_{t-1}) \right) .
\end{align*} 

The FF-BS is an exact sampling algorithm and in principle, could be applied to FHMMs. However, this becomes infeasible even for a moderate number of latent sequences $K$. This is due to the state space growing exponentially with $K$. As the full FF-BS has complexity $O(2^{2K} T)$, a computationally cheaper approach is needed, however this comes at the expense of sampling efficiency.

One option is to sample each row of $\boldX$ conditional on the rest, using the FF-BS. Then each of the updates has a state space of size 2 and the FF-BS steps are inexpensive. However, in this conditional scheme most of the sequences are fixed and thus it is difficult for the sampler to explore the space well. A more general version of this would update a small subset of chains jointly at a higher computational cost, which can still get trapped in local modes. 

An alternative idea referred to as Hamming Ball sampling has been suggested by \citet{titsias2014hamming, titsias2017hamming}, which adaptively truncates the space via an auxiliary variable scheme. Unlike the conditional Gibbs updates, it does not restrict parts of $\boldX$ to be fixed during sampling. Even though it can be less prone to get stuck, for a moderate value of $K$ it may still not explore the whole posterior space.


This problem can be alleviated by ensemble MCMC methods which combine ideas from simulated annealing \citep{kirkpatrick1983optimization} and genetic algorithms \citep{holland1992adaptation}. 
One such example is parallel tempering \citep{geyercomputing}.
Instead of running a single chain targeting the posterior, one introduces an ensemble of chains and assigns a temperature to each chain so that every chain would be targeting a tempered version of the posterior. Tempered targets are less peaked and therefore higher temperature chains in the ensemble explore the space well and do not get stuck. The key question becomes how to efficiently exchange information between the chains.

In this paper, we propose a novel ensemble MCMC method which provides an auxiliary variable construction to exchange information between chains. This is a general MCMC method, but our main focus is on improving existing poorly mixing samplers for sequence-type data. Specifically we consider the application to Factorial Hidden Markov Models. 
We demonstrate the practical utility of our augmented ensemble scheme in a series of numerical experiments, covering a toy sampling problem as well as inference for FHMMs. The latter involves a simulation study as well as a challenging cancer genomics application.

\section{Augmented ensemble MCMC} 


Monte Carlo-based Bayesian inference \citep{andrieu2003introduction} for complex high-dimensional posterior distributions is a challenging problem as efficient samplers need to be able to move across irregular landscapes that may contain many local modes \citep{gilks1996strategies, frellsen2016bayesian, betancourt2017conceptual}. Commonly adopted sampling approaches can explore the space very slowly or become confined to regions around local modes.

Ensemble MCMC (also known as population-based MCMC, or evolutionary Monte Carlo) methods \citep{jasra2007population,neal2011mcmc,shestopaloff2014efficient} can alleviate this problem. This is achieved by introducing an ensemble of MCMC chains and then exchanging information between the chains. Next, we review standard ensemble MCMC approaches and proposal mechanisms that are used to exchange information. Then, we introduce our augmented Gibbs sampler. 

\subsection{Standard ensemble sampling methods \label{sec:standardEnsemble}}

Suppose our goal is to sample from a density $\pi$. Instead of sampling $\boldx \sim \pi(\cdot)$, ensemble MCMC introduces an extended product space $(\boldx_1, \ldots, \boldx_K)$ with a new target density $\pi^*$ defined as 
$
	\pi^*(\boldx_1, \ldots, \boldx_K) = \prod_{k=1}^K \pi_k (\boldx_k), 
$
where $\pi_{k} = \pi$ for at least one index. Here we focus on parallel tempering, which introduces a temperature ladder $T_1 < \ldots < T_K$ and associates a temperature with each chain. Denoting the inverse temperature $\beta_k = 1/T_k$, we define the tempered targets $\pi_k(\boldx_k) := \pi(\boldx_k)^{\beta_k}$. The idea is that high temperature chains can readily explore the space since the density is flattened by the power transformation, whereas the chain containing the true target density with $T_1 = 1.0$ only samples locally and precisely from the target. 
Each chain is updated independently, with occasional information exchange between the chains so that more substantial movement in the higher temperature chains can filter down to the slower moving low temperature chains.

One approach to exchanging information is to propose swapping states (``swap'' move) between chains of consecutive temperatures and then performing an accept/reject operation according to the Metropolis-Hastings ratio \citep{geyercomputing, earl2005parallel}. However, a global move like this is unlikely to be accepted in a high-dimensional sampling setting. 

More elaborate approaches can create proposals using genetic algorithms \citep{liang2000evolutionary}, by proposing certain moves between chains which again requires accepting/rejecting based on the Metropolis-Hastings framework. 
One such proposal scheme is a one-point crossover move, illustrated as follows:
\[
\begin{aligned}
(x_1, \ldots, x_t, x_{t+1}, \ldots, x_T) \\
(y_1, \ldots, y_t, y_{t+1}, \ldots, y_T) \\
\end{aligned}
\implies
\begin{aligned}
(y_1, \ldots, y_t, x_{t+1}, \ldots, x_T) \\
(x_1, \ldots, x_t, y_{t+1}, \ldots, y_T) \\
\end{aligned}
\]
where the crossover point $t$ could for example be chosen uniformly $t \in \{1,\ldots,T\}$. This is most natural for sequential models where there is dependency between consecutive $x_t$ and $x_{t+1}$. For high-dimensional sequences this is more appealing than a swap move due to being more local and thus leading to higher chance of acceptance. One can similarly construct a two-point crossover move. 
However, the accept/reject procedure can be inefficient and very sensitive to \emph{both} the choice of the temperature ladder and algorithmic parameter tuning. Our work seeks to address the latter issue by using an auxiliary variable augmentation that produces a Gibbs sampling scheme.

\setlength{\abovedisplayskip}{3pt}
\setlength{\belowdisplayskip}{3pt}

\subsection{Gibbs sampling using auxiliary variables \label{sec:auxRejfree}}

Now, consider the target in the product space $\pi^*$. Suppose that during MCMC we would like to exchange information between a pair of chains $\pi_i(\boldx_i)$ and $\pi_j(\boldx_j)$ where  $\boldx_i$ and $\boldx_j$ are $T$-dimensional vectors that indicate the current states of these chains. Here we describe an auxiliary variable move, which uses the idea of a one-point crossovers and leads to a Gibbs update for a two-point crossover.

We introduce two auxiliary variables $\boldu$ and $\boldv$, that live in the same space as $\boldx_i$ and $\boldx_j$, drawn from an auxiliary distribution $p(\boldu, \boldv | \boldx_i, \boldx_j)$. Without loss of generality we assume that this auxiliary distribution is uniform over all possible one-point crossovers between $\boldx_i$ and $\boldx_j$. 

We also introduce the set $\textsc{cr}(\boldx, \boldy)$ to denote all $T$ crossovers between the vectors $\boldx$ and $\boldy$. The auxiliary distribution $p(\boldu, \boldv | \boldx_i, \boldx_j)$ is precisely a uniform distribution over all pairs $(\boldu, \boldv) \in \textsc{cr}(\boldx_i, \boldx_j)$. This distribution is also symmetric, i.e. $p(\boldu, \boldv | \boldx_i, \boldx_j) = p(\boldx_i, \boldx_j | \boldu, \boldv)$.

Using the auxiliary variables we can exchange information between $\boldx_i$ and $\boldx_j$ through the intermediate step of sampling the auxiliary variables $(\boldu,\boldv)$ based on the following two-step Gibbs procedure: 
\begin{enumerate}
\item Generate $(\boldu, \boldv) \sim p(\boldu, \boldv | \boldx_i, \boldx_j)$
\item Generate $(\boldx_i, \boldx_j) \sim p(\boldx_i, \boldx_j | \ee)$, where
\begin{align*}
p(\boldx_i, \boldx_j | \ee) &= \frac{1}{Z} \pi_i(\boldx_i) \pi_j(\boldx_j)  p(\boldu, \boldv | \boldx_i, \boldx_j) \\
&= \frac{1}{Z} \pi_i(\boldx_i) \pi_j(\boldx_j)  p(\boldx_i, \boldx_j | \boldu, \boldv) \\
&= \frac{1}{Z} \pi_i(\boldx_i) \pi_j(\boldx_j) I((\boldx_i, \boldx_j) \in \textsc{cr} (\boldu, \boldv)) 
\end{align*}
\end{enumerate}
where the normalising constant $Z = Z(\boldu, \boldv)$ is
$Z(\boldu, \boldv) = \sum_{(\boldy_i, \boldy_j) \in \textsc{cr}(\boldu, \boldv)} \pi_i(\boldy_i) \pi_j(\boldy_j).$

The first step of the above procedure selects a random crossovered pair $(\boldu, \boldv)$, while the second step 
conditions on this selected pair and jointly samples $(\boldx_i,\boldx_j)$ from the exact conditional posterior 
distribution that takes into account the information coming from the actual chains $\pi_i$ and $\pi_j$. 

\begin{figure}[t]
\vskip -0.5em
\begin{center}
\centerline{\includegraphics[width=\columnwidth]{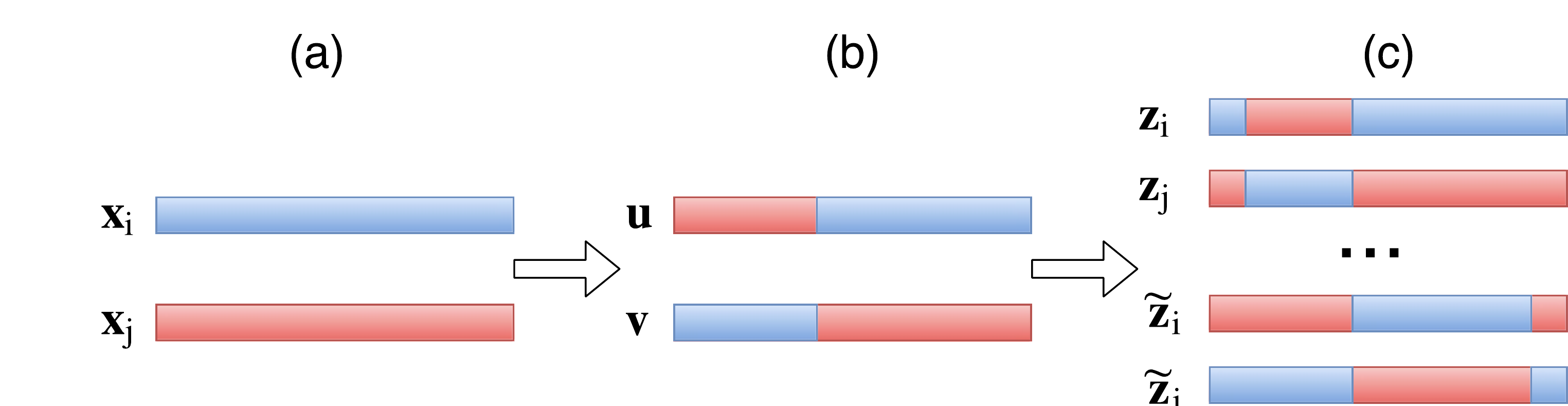}}
\caption{Schematic overview of the auxiliary variable crossover move. (a) We start with two sequences $\boldx_i$ and $\boldx_j$. (b) Now we construct auxiliary variables $\boldu, \boldv$ by applying a uniform one-point crossover to $\boldx_i, \boldx_j$. (c) Next, we consider all possible crossovers of $\boldu, \boldv$, and according to probabilities $\pi_i(\boldz_i) \pi_j(\boldz_j)$, we accept one of these configurations as the new value of $\boldx_i, \boldx_j$.}
\label{fig:schema_crossover}
\end{center}
\vskip -0.2in
\end{figure} 

Since the above is a Gibbs operation it leads to new state vectors for the chains $\pi_i$ and $\pi_j$ that are always accepted. To prove this explicitly we compute the effective marginal proposal and show that the corresponding Metropolis-Hastings acceptance probability is always one.

Given the current states $(\boldx_i, \boldx_j)$, we denote the proposed states by $(\boldz_i, \boldz_j)$ 
and the marginal proposal distribution by $Q(\boldz_i, \boldz_j | \boldx_i, \boldx_j)$. This proposal, defined by the above 
two-step Gibbs procedure, is a mixture: 
\begin{align*}
&Q(\boldz_i, \boldz_j | \boldx_i, \boldx_j) = \\
&= \iint \frac{1}{Z} \pi_i(\boldz_i) \pi_j(\boldz_j) p(\boldz_i, \boldz_j | \boldu, \boldv) p(\boldu, \boldv | \boldx_i, \boldx_j) d\boldu d\boldv \\
&= \pi_i(\boldz_i) \pi_j(\boldz_j) \iint \frac{1}{Z} p(\boldz_i, \boldz_j | \boldu, \boldv) p(\boldx_i, \boldx_j | \boldu, \boldv) d\boldu d\boldv \\
&= \pi_i(\boldz_i) \pi_j(\boldz_j) H(\boldz_i, \boldz_j | \boldx_i, \boldx_j).
\end{align*}

The Metropolis-Hastings acceptance probability under this proposal is
\begin{align*}
\alpha 
&= \frac{\pi_i(\boldz_i) \pi_j(\boldz_j) Q(\boldx_i, \boldx_j | \boldz_i, \boldz_j)}{\pi_i(\boldx_i) \pi_j(\boldx_j) Q(\boldz_i, \boldz_j | \boldx_i, \boldx_j)} \\
&= \frac{\pi_i(\boldz_i) \pi_j(\boldz_j) \pi_i(\boldx_i) \pi_j(\boldx_j) H(\boldx_i, \boldx_j | \boldz_i, \boldz_j)}{\pi_i(\boldx_i) \pi_j(\boldx_j) \pi_i(\boldz_i) \pi_j(\boldz_j) H(\boldz_i, \boldz_j | \boldx_i, \boldx_j)}.
\end{align*}

Since $H(\boldz_i, \boldz_j | \boldx_i, \boldx_j) = H(\boldx_i, \boldx_j | \boldz_i, \boldz_j)$ due to symmetry, all terms cancel out and $\alpha = 1$. So our proposal will be always accepted. 

To simulate from $Q(\boldz_i, \boldz_j | \boldx_i, \boldx_j)$ in practice, we can use its mixture representation above, i.e. first generate auxiliary variables $p(\boldu, \boldv | \boldx_i, \boldx_j)$ and then conditional on those, generate the new value from $p(\boldz_i, \boldz_j | \boldu, \boldv)$. We note that even though both of these steps are implemented as one-point crossovers, the overall proposal can lead to a two-point crossover as illustrated in Figure~\ref{fig:schema_crossover}.

Specifically, to implement this, first we sample $(\boldu, \boldv)$ uniformly from the set $\textsc{cr}(\boldx_i, \boldx_j)$. Now, conditional on the obtained $(\boldu, \boldv)$, let us denote the crossover of $\boldu$ and $\boldv$ at point $t$ by $(\boldz_i^{(t)}, \boldz_j^{(t)})$. The second step is to iterate 
over $t \in \{1, \ldots, T\}$ and compute quantities $a_t := \pi_i(\boldz_i^{(t)}) \pi_j(\boldz_j^{(t)})$. The pair $(\boldz_i^{(t)}, \boldz_j^{(t)})$ will be accepted as the new value of $(\boldx_i, \boldx_j)$ with probability proportional to $a_t$. 

A further extension of the above procedure is obtained by modifying the auxiliary distribution 
$p(\boldu, \boldv |\boldx_i, \boldx_j)$ to become uniform over the union of the 
sets $\textsc{cr}(\boldx_i, \boldx_j)$ and $\textsc{cr}(\boldx_j, \boldx_i)$ since, 
due to the deterministic ordering, the crossovers between $\boldx_i$ with $\boldx_j$ 
and the reverse crossovers between $\boldx_j$ with $\boldx_i$ are not identical. The auxiliary
distribution $p(\boldu, \boldv |\boldx_i, \boldx_j)$ still remains symmetric and all above 
properties hold unchanged. The only difference is that now we are considering 
$2 T$ crossovers and in order to sample from $p(\boldu, \boldv |\boldx_i, \boldx_j)$ 
we need first to flip a coin to decide the order of $\boldx_i$ and $\boldx_j$. 
Complete pseudocode of the whole procedure is given in Supplementary.

The above sampling scheme is general and it can be applied to arbitrary 
MCMC inference problems involving both continuous and discrete variables. In the next 
section we apply the proposed method to a challenging inference problem in Factorial HMMs (FHMMs). 

\section{Application to FHMMs \label{sec:ensembleFHMM}}

Here, we apply the augmented ensemble scheme to FHMMs in order to improve on existing poorly mixing samplers. We achieve this via an ensemble of chains over suitably defined tempered posteriors. For a latent variable model, one can either temper the whole joint distribution or just the emission likelihood. We chose the latter, so the target posterior of interest becomes
$$
\pi_k(\boldX) := p_k(\boldX | y_{1:T}) \propto p(\boldX) p(y_{1:T} | \boldX)^{\beta_k}
$$
where $\boldX$ is a $K \times T$ binary matrix. As the \crossovers was originally defined on vectors, there are multiple ways to extend this to matrices. One can perform crossovers on either rows or columns of a matrix, potentially considering a subset of those. Here we have decided to focus on a crossover move defined on the \emph{rows} of $\boldX$, specifically on all $K$ rows of $\boldX$. 
 
The core computational step of the algorithm is to compute quantities $a_t$ for all crossover points $t$. We show that these can be computed recursively in an efficient way. 
Let $\boldU$ and $\boldV$ be the current states of the auxiliary matrices for chains $i$ and $j$. Comparing their crossovers at two consecutive points $t-1$ and $t$, denoted by $(\boldZ^{(i)}_{t-1}, \boldZ^{(j)}_{t-1})$ and $(\boldZ^{(i)}_{t}, \boldZ^{(j)}_{t})$, we note that these can differ just in column $t$:
\begin{align*}
\boldZ^{(i)}_{t-1} := (\boldv_1, \ldots, \boldv_{t-1}, \boldu_t, \boldu_{t+1}, \ldots, \boldu_T), \\
\boldZ^{(i)}_{t} := (\boldv_1, \ldots, \boldv_{t-1}, \boldv_t, \boldu_{t+1}, \ldots, \boldu_T).
\end{align*}
As a result, the values $a_t = \pi_i(\boldZ_t^{(i)}) \pi_j(\boldZ_t^{(j)})$ can be computed recursively. Indeed, given the previous value of $\pi_i(\boldZ_{t-1}^{(i)})$, we can compute $\pi_i(\boldZ_{t}^{(i)})$ by accounting for the following two cases: first, change in emission likelihood from $p(y_t | \boldu_t)^{\beta_i}$ to $p(y_t | \boldv_t)^{\beta_i}$, and second, change in the transitions from $\boldv_{t-1} \rightarrow \boldu_t \rightarrow \boldu_{t+1}$ to $\boldv_{t-1} \rightarrow \boldv_t \rightarrow \boldu_{t+1}$. 

By denoting the overall transition probability $p(\boldu_{t+1} | \boldu_t)$ for chain $i$ by $A^{(i)}(\boldu_t, \boldu_{t+1})$, 
we can express $\pi_i(\boldZ^{(i)}_{t})$ in terms of $\pi_i(\boldZ^{(i)}_{t-1})$ as follows
$
\pi_i(\boldZ^{(i)}_{t}) = \pi_i(\boldZ^{(i)}_{t-1}) \cdot  c^{(i)}_t 
$
where 
\[
c^{(i)}_t := \frac{ A^{(i)}(\boldv_{t-1}, \boldv_t) A^{(i)}(\boldv_{t}, \boldu_{t+1}) }
{ A^{(i)}(\boldv_{t-1}, \boldu_t) A^{(i)}(\boldu_{t}, \boldu_{t+1}) } \cdot
\frac{p(y_t | \boldv_t )^{\beta_i}}{p(y_t | \boldu_t )^{\beta_i}} .
\]
Now we can compute the quantities $a_t$ recursively as follows
$
a_t = a_{t-1} \cdot c^{(i)}_t \cdot c^{(j)}_t .
$
As the values of $a_t$ can be normalised to sum to one, we can arbitrarily fix the reference value $a_0 \gets 1$. The computation of every correction term $a_t$ is of the complexity $O(K)$, and the overall cost for all $a_t$ values is $O(KT)$, being relatively cheap. As we typically need to perform the crossover moves only occasionally, the \crossovers provides a way to improve the poorly mixing samplers for FHMMs at a small extra computational cost.

\section{Experiments}

First, we demonstrate the proposed sampling method on a multimodal toy inference problem. Then, we focus on Bayesian inference for FHMMs: we compare various samplers in a simulation study and then consider a challenging tumor deconvolution example.
In both experiments, we compare a standard single-chain sampling technique (a Gibbs sampler or the Hamming Ball sampler) with the respective ensemble versions. 

For ensemble samplers, we compare our proposed augmentation scheme (``augmented crossover'') with two additional baseline exchange moves: the standard swap move (``swap'') and a uniformly chosen crossover (``random cr'') within the accept-reject Metropolis-Hastings framework. In all experiments, we run an ensemble of two MCMC chains, with temperatures $T_1 = 1.0$ and $T_2 = 5.0$, carrying out an exchange move every 10-th iteration. 

\subsection{Toy example \label{sec:toy}}

\begin{figure}[!t]
\centering
\includegraphics[width=\columnwidth]{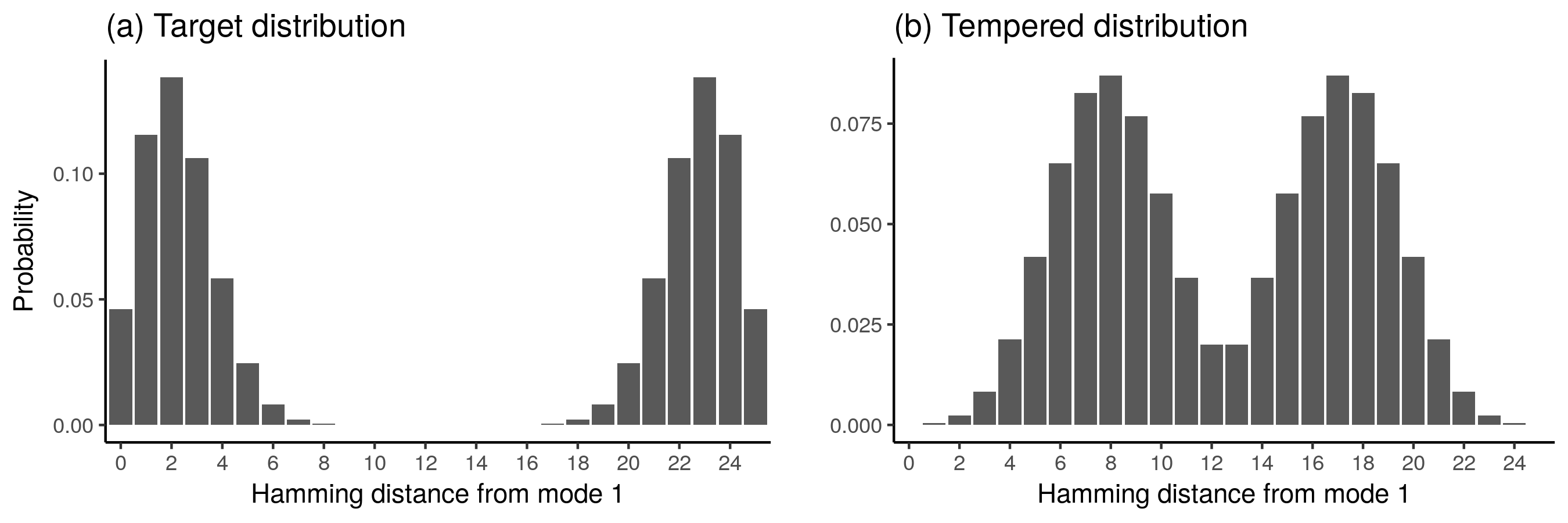}
\caption{The bimodal within-block probability distribution as defined in eq.~\eqref{eq:toy} for a binary sequence of length 25 shown in (a) and its tempered version in (b).}
\label{fig:toy_distr}
\vspace*{\floatsep}
\includegraphics[width=0.5\columnwidth]{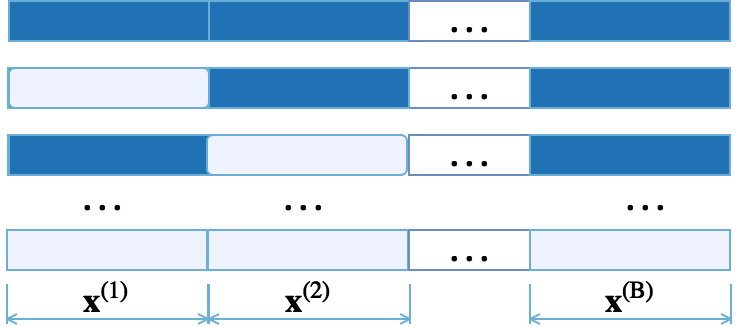}
\caption{Multiple modes of the distribution of $(\boldx^{(1)}, \ldots, \boldx^{(B)})$, colour coding: dark = 1, light = 0.}
\label{fig:toy_multimodal}
\vspace*{\floatsep}
\includegraphics[width=\columnwidth]{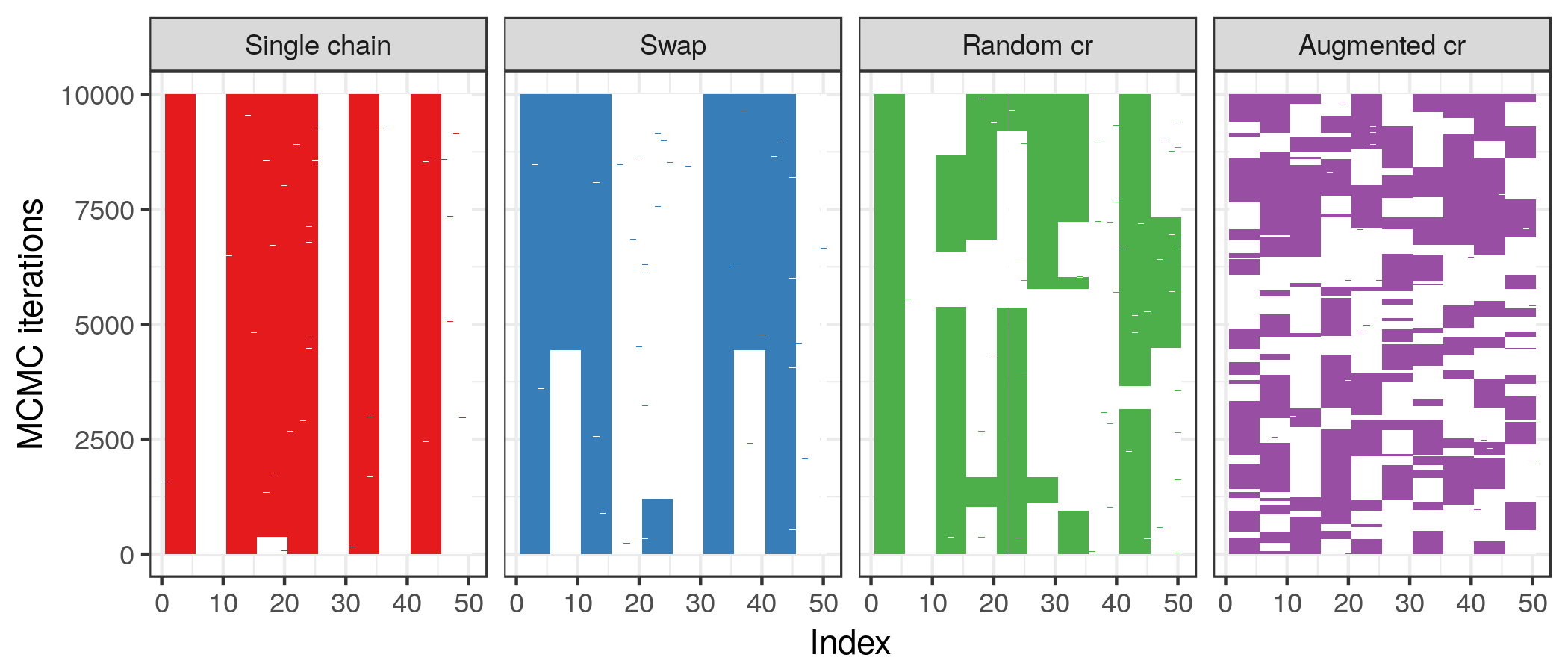}
\caption{Heatmaps representing the trace plots of $\boldx$ for the experiment with $B=10$ blocks, running a single chain Gibbs sampler (first panel), and its ensemble versions with various exchange moves: swap, random crossover, augmented crossover (in four panels). For each MCMC iteration, the elements of $\boldx$ have been colour coded: dark = 1, light = 0.}
\label{fig:b10}
\vspace*{\floatsep}
\includegraphics[width=\columnwidth]{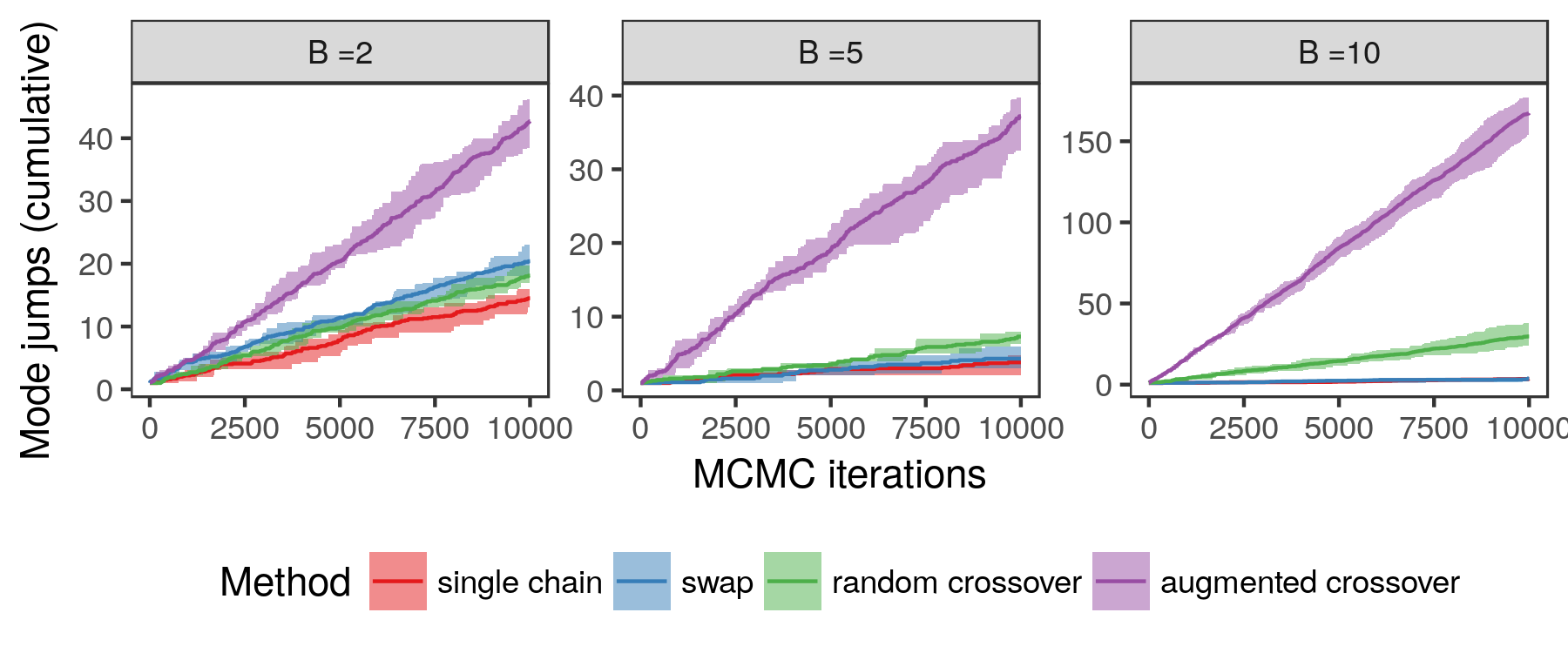}
\caption{The cumulative number of jumps between modes ($y$-axis) over MCMC iterations ($x$-axis) for various experiments (block sizes $B \in \{2, 5, 10\}$) on average (coloured lines) together with 25\% and 75\% quantiles (shaded areas) over 10 repeated runs.}
\label{fig:toy_njumps}
\end{figure}

We consider the following multimodal toy sampling problem, where the target distribution is binary and has multiple separated modes. Specifically, we fix the dimensionality $T=50$ and divide the sequence $\boldx$ into $B$ contiguous blocks as follows $(\boldx^{(1)}, \ldots, \boldx^{(B)})$. In each of the blocks, we define a bimodal distribution, having two peaked modes $\boldx^{\text{mode}_1} := (1, 1, \ldots, 1)$ and $\boldx^{\text{mode}_2} := (0, 0, \ldots, 0)$, such that the probability of any binary vector $\boldx^{(j)}$ in block $j$ is given by
\begin{equation} \label{eq:toy}
p(\boldx^{(j)}) \propto \alpha_j^{\min(\dist(\boldx^{(j)},\, x^{\text{mode}_1}),\; \dist(\boldx^{(j)},\, \boldx^{\text{mode}_2}))}
\end{equation}
where $\dist(\cdot, \cdot)$ denotes the Hamming distance between two binary vectors and $\alpha_j$ is a block-specific parameter which controls how peaked the modes are. As a result, the further we go from the modes (in terms of Hamming distance), the less likely we are to observe that state. This has been illustrated in Figure~\ref{fig:toy_distr}. 


We extend the above to define the joint $p(\boldx)$ factorising over the blocks as follows
\[
p(\boldx) \propto \prod_{j=1}^B p(\boldx^{(j)}) = \prod_{j=1}^B \alpha_j^{\min(\dist(\boldx^{(j)},\, \boldx^{\text{mode}_1}),\; \dist(\boldx^{(j)},\, \boldx^{\text{mode}_2}))}
\]
Within each block, the probability of a given state depends on its distance to the closest mode. 
This construction induces strong within-block dependencies. By varying the number of blocks within a sequence of fixed length, we can interpolate between a strong global correlation and local dependencies with a highly multimodal structure. 
The total number of modes for this distribution is $2^B$, as illustrated in Figure~\ref{fig:toy_multimodal}. 

In our experiments, we vary $B \in \{2, 5, 10\}$, resulting in distributions having $2^2, 2^5, 2^{10}$ modes. We generate $\alpha_j \in U(\{0.01, 0.02, \ldots, 0.05\})$. All samplers are initialised from the same value (one of the modes) and run for 10,000 iterations.

The resulting traces of $\boldx$ have been shown as heatmaps in Figure~\ref{fig:b10} for $B=10$ (see Supplementary Figures for $B=2$ and $B=5$). 
As a summary statistic, we have shown the cumulative number of jumps between modes over repeated experiments in Figure~\ref{fig:toy_njumps} . 

In all scenarios, the single chain Gibbs sampler expectedly struggles to escape the mode from which it was initialised, with ensemble methods better at moving between modes. 
For strong global correlations (corresponding to small $B$ values), the baseline exchange moves ``swap'' and ``random crossover'' are reasonably efficient, though still result in a smaller number of mode jumps than the ``augmented crossover''. 

Now when increasing $B$, the dependency structure becomes more local, resulting in a much more multimodal sampling landscape. For $B=10$, the simple ``swap'' and ``random crossover'' moves struggle to accept any proposals at all and the benefit of our augmentation scheme becomes clear. In this highly multimodal setting with $2^{10}$ modes, the total number of modes visited by our ``augmented crossover'' (average 144) is much higher than for the ``swap'' (3) and ``random crossover'' (27) moves.

\subsection{Tumor deconvolution example \label{sec:tumor}}

The following example is motivated by an application in cancer genomics. Certain mutations in the cancer genome result in a loss of DNA integrity leading to copy number alterations due to the duplication or loss of certain DNA regions. Tumor samples consist of heterogeneous cell subpopulations and it is of interest to identify the subpopulations to study their phylogeny and gain insight into the clonal evolution \citep{ha2014titan, gao2016punctuated}. However, as DNA sequencing of bulk tissue samples produces aggregate data over all constituent cell subpopulations, the observed sequencing read counts must be deconvolved to reveal the underlying latent genetic architecture. 

\begin{figure}[!t]
\begin{center}
\includegraphics[width=\columnwidth]{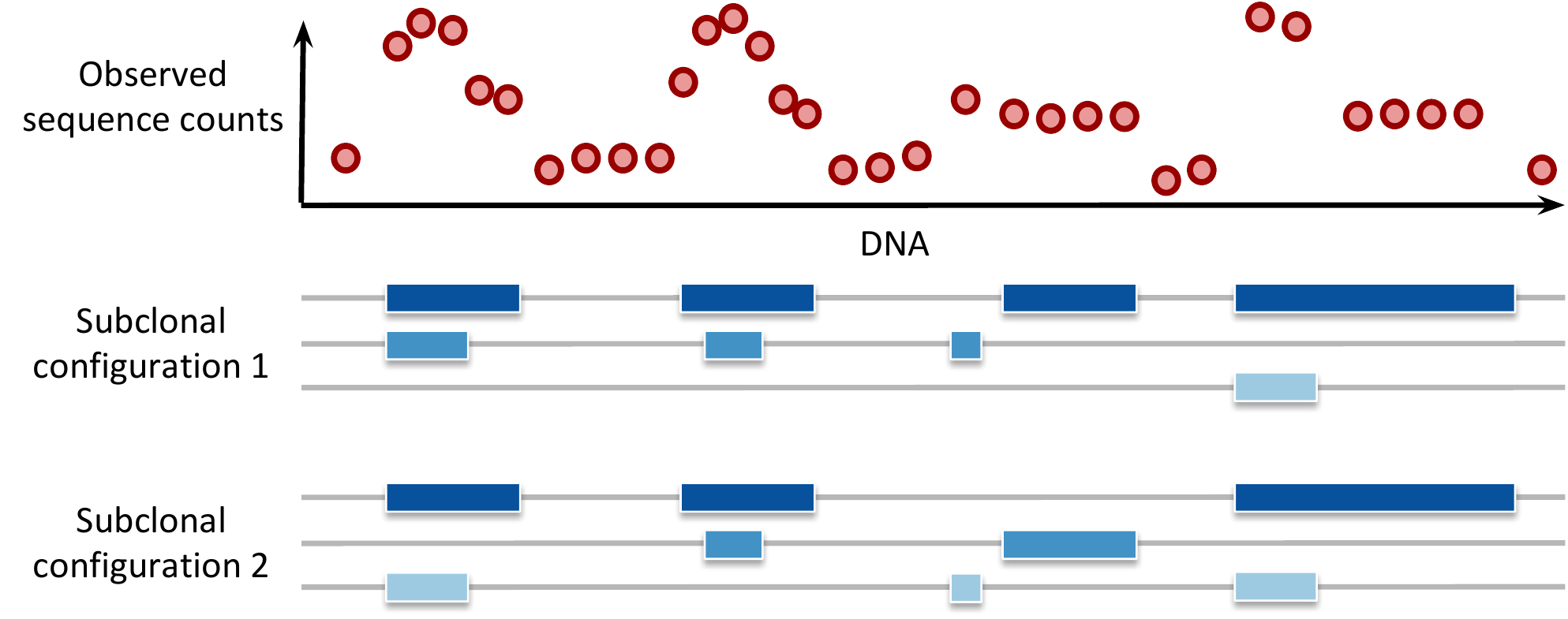}
\caption{Existence of multiple subclonal configurations, both consisting of $K=3$ subpopulations, which are indistinguishable when sequence counts (in top panel) are observed as aggregate over subpopulations. }
\label{fig:subclonal}
\end{center}
\end{figure}

The additive Factorial HMM is a natural model to consider where each latent chain corresponds to a putative cell subpopulation. However, it is important that the exploration of the state space of the latent chains allows us to identify the different subpopulation configurations that are compatible with the observed sequencing data since there maybe a number of plausible possibilities. This is illustrated in Figure~\ref{fig:subclonal}. A poorly mixing sampler which is exploring only one of the possible latent explanations could lead to misleading conclusions regarding the subclonal architecture of a tumor. We wanted to examine if the ensemble scheme we propose could provide a more effective means of posterior sampling.

\subsubsection{Simulation study}

Lets consider the emission model
$
y_t | \boldx_t, \boldw, h \sim \N \left (h \sum_{k=1}^K w_k x_{k, t}, \; \sigma^2 \right )
$
where $y_t$ denote the observed sequence read counts at a locus $t$ and $h$ is the expected sequencing depth. Each $w_k$ corresponds to the fraction of $k$-th subpopulation ($w_k \ge 0$, $\sum_k w_k = 1$) whose mutation profile is given by the $k$-th row of $\boldX$. Here $x_{k,t} \in \{0, 1\}$ denotes whether the $k$-th population has a copy number alteration at position $t$ or not. 

Note that this is not a complete model of real-world sequencing data but a simplified version to demonstrate the utility of the proposed ensemble MCMC methods. The results presented here should extend to the more complex cases. Further work to construct a sufficiently complex model to capture the variations within real sequencing data, such as single nucleotide polymorphisms, is beyond the scope of this paper and will be developed in future work. 

First, we investigated the performance of sampling schemes for FHMMs in the presence of multimodality in a controlled setting. We generated observations from the emission distribution with $K=3$ with weights such that $w_1 + w_2 \approx w_3$. As a result, data generating scenarios $\boldx_t = (1, 1, 0)$ and $\boldx_t = (0, 0, 1)$ are both plausible underlying latent explanations.

\begin{figure}[!ht]
\begin{center}
\includegraphics[width=\columnwidth]{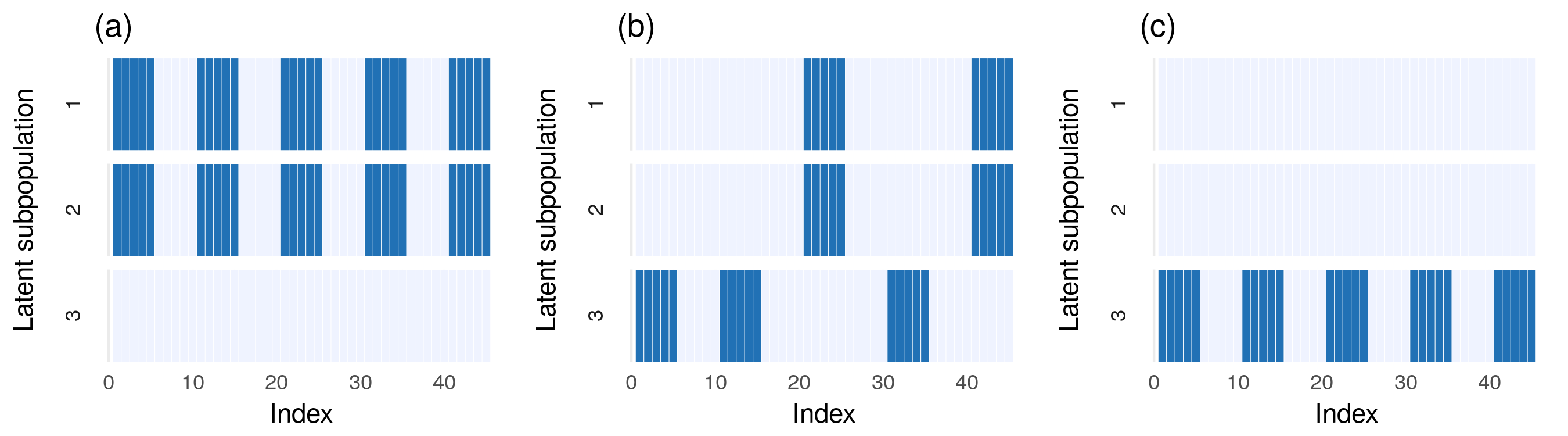}
\vspace{-5mm}
\caption{Small illustration of three possible modes for the $\boldX$ matrix used in the simulation study. }
\label{fig:generatedX}
\end{center}
\end{figure}

For data generation, we used a latent $\boldX$ matrix having a block structure of columns $(1, 1, 0)$ followed by a block of $(0, 0, 0)$, as illustrated in Figure~\ref{fig:generatedX}(a), but using altogether 20 blocks. We fixed $h=15$, $\boldw = (0.2+\varepsilon, 0.3+\varepsilon, 0.5-2\varepsilon)$ with $\varepsilon = 0.01$ and $\sigma^2=1$. Each of these blocks has two modes, but due to the structured FHMM prior on $\boldX$, the mode $(0, 0, 1)$ corresponds to a slightly higher log-posterior value. For example, the three examples provided in Figure~\ref{fig:generatedX} are ordered in terms of posterior probability (c) $>$ (b) $>$ (a). 


\begin{figure*}
\centering
\includegraphics[width=\textwidth]{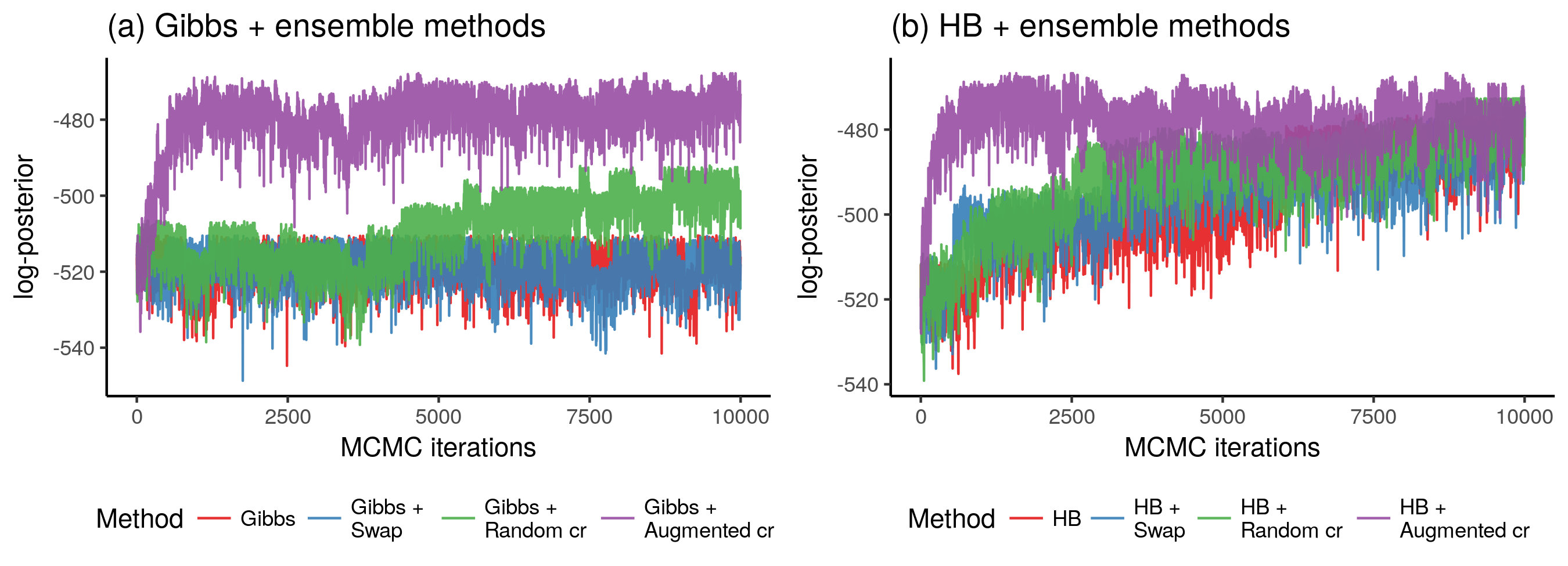}
\includegraphics[width=\textwidth]{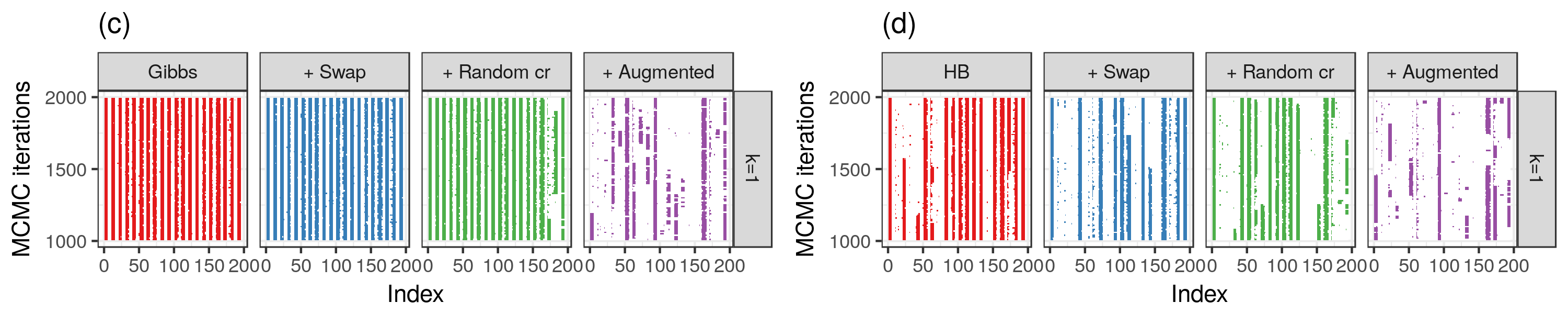}
\includegraphics[width=\textwidth]{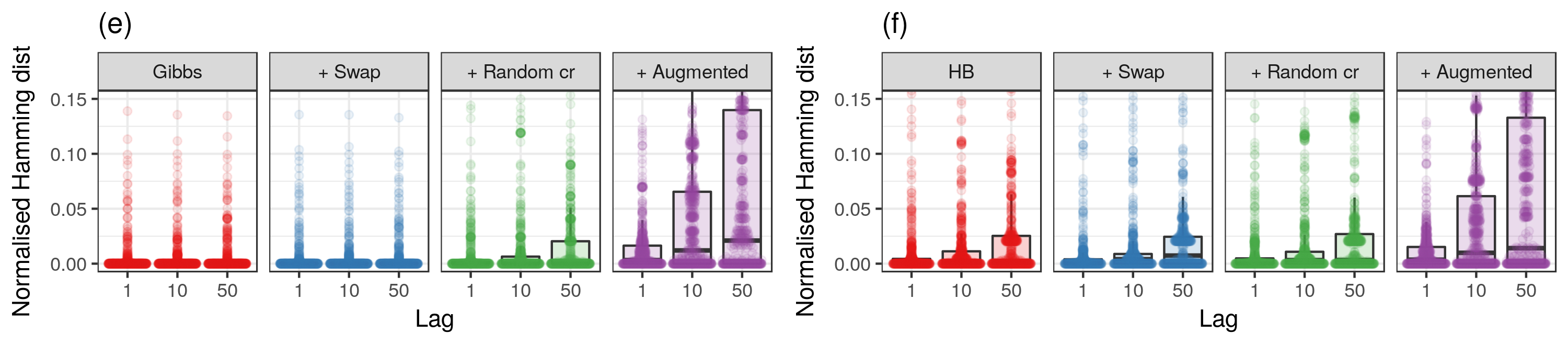}
\vspace{-3mm}
\caption{Simulation study comparing sampling techniques for FHMMs: ensemble versions of the Gibbs sampler (left column) and of the Hamming Ball sampler (right column). (a, b) Traces of log-posterior for the single chain sampler (``Gibbs'', ``HB'') and three ensemble versions. (c, d) Heatmaps showing the traces for the first row ($k=1$) of $\boldX$ (colour coded: dark = 1, light = 0), zoomed in to MCMC iterations 1000 - 2000 ($y$-axis). (e,f) Distribution of the normalised Hamming distance: boxplots in the background, overlaid with individual values) for various lags 1, 10, 50 ($x$-axis).}
\label{fig:toy_FHMM}
\end{figure*}

For inference in FHMMs, we considered two single chain samplers for $\boldX$: one-row updates conditional on the rest (``Gibbs''), and the Hamming Ball sampler (``HB''). We then considered ensemble versions of both of these samplers, as shown in Figure~\ref{fig:toy_FHMM} (left column for ``Gibbs'' and right column for ``HB''). 
All chains were initialised from the mode with $x_{3, t} = 0$, i.e. mode (a) in Figure~\ref{fig:generatedX}, and were ran for 10~000 iterations. Exchange moves were carried out every 10th iteration. 

\begin{figure*}
\centering
\includegraphics[width=\textwidth]{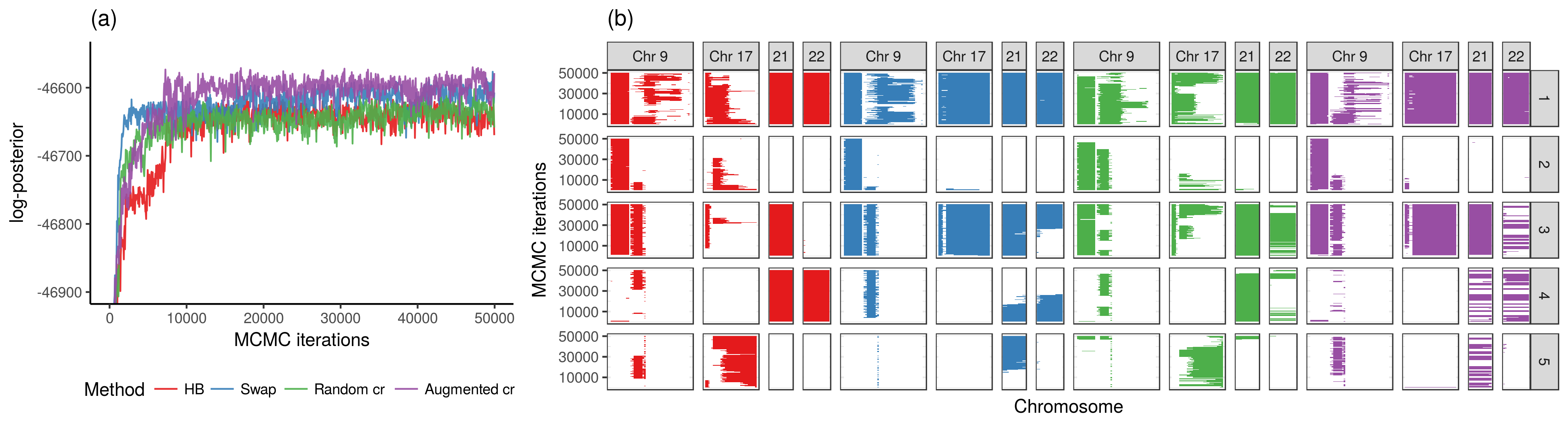}
\vspace{-5mm}
\caption{FHMM fitted to real sequencing data, using HB$(r=3)$ single-chain sampler and the corresponding ensemble samplers (``swap'', ``random cr'', ``augmented cr''). (a) Traces of log-posterior ($y$-axis), and (b) traces of $\boldX$ over MCMC iterations for each $5$ rows of $\boldX$ (row panels 1 -- 5), with the genomic coordinates ($x$-axis) zoomed in to selected chromosomes.}
\label{fig:real_traces}
\end{figure*}

For ``Gibbs'', the single chain sampler and the ``swap'' ensemble have not moved from the initialisation, the ``random cr'' ensemble scheme shows some improvement, but the ``augmented cr'' has quickly moved towards values of $\boldX$ with higher posterior probability (see Figure~\ref{fig:toy_FHMM}(a)). It also exhibits much better mixing, as seen from the traces of the first row of $\boldX$, i.e. traces of $x_{1, 1:T}$ shown in Figure~\ref{fig:toy_FHMM}(c). We note that $x_{1, t} = 0$ values correspond to the more probable mode. 

As a single chain sampler, ``HB'' quickly achieves higher log-posterior values than ``Gibbs''. Therefore, for ``HB'' the gain from ``swap'' and ``random cr'' ensemble techniques is relatively smaller, but still the ``augmented cr'' has quickly moved towards higher log-posterior values. 

To quantify mixing on binary state spaces, we have calculated the Hamming distance between $\boldX^{(t)}$ and $\boldX^{(t + \text{lag})}$ for various lag values $\{1, 10, 50\}$, normalised by $\text{dim}(\boldX)$. Panels (e, f) show the distribution of these summary statistics, confirming that the augmented crossover scheme reduces notably the dependence between consecutive samples of $\boldX$.

\begin{table}[!t]
\captionof{table}{Computation times in seconds for the simulation study (two chains, 10~000 iterations). 
}
\centering
\begin{tabular}{rrrr}
  \hline
HB & swap & random cr & augmented cr \\ 
130 & 132 & 133 & 135 \\ 
   \hline
Gibbs & swap & random cr & augmented cr \\ 
213 & 216 & 214 & 218 \\ 
   \hline
\end{tabular}
\label{tab:comptimes}
\vspace{-1em}
\end{table}

We have shown above that the complexity of augmented crossover scheme is linear $O(KT)$, which is also the case for the ``swap'' and ``random cr'' moves. To explore the respective costs in practice, we measured the total computation time for our Rcpp implementation. To establish the baseline cost of running a two-chain ensemble without any exchange moves in a sequential implementation, we indicate this baseline time in the first column (``Gibbs'' and ``HB'') of Table~\ref{tab:comptimes}. We note that this could be halved by a parallel implementation. The extra cost for all exchange moves are relatively small. Even though the extra time for the ``swap'' and ``random cr'' schemes is just slightly smaller than for ``augmented cr'', this is a small price to pay for an improvement in mixing, especially compared to the high baseline cost of running an FHMM sampler.


\subsubsection{Tumor data analysis}

Next we consider whole-genome tumor sequencing data for bladder cancer \citep{cazier2014whole}. 
To illustrate the utility of our sampling approach, we used data from one patient (patient ID: 451) and took a thinned sample of 10,877 loci. We placed a vague Gaussian prior on the expected sequencing depth, $h \sim \N(\mu_h, \sigma^2_h)$ with $\mu_h = 180, \sigma_h = 30$, and integrated out $h$, resulting in the marginal likelihood
\begin{align*}
y_t | \boldx_t, \boldw \sim \N \left (\mu_h \sum_{k=1}^K w_k x_{k, t}, \; \sigma^2 + \sigma^2_h \left( \sum_{k=1}^K w_k x_{k, t} \right)^2 \right ) .
\end{align*}
Here each row of $\boldX$ corresponds to a single chromosome and the binary state indicates whether a copy of that DNA region exists or not. We fixed $K=6$, where one of the latent sequences is always fixed to 1, representing a baseline, unaltered chromosome. 
We used a Hamming Ball Sampler with radius $r = 3$ as a single chain sampler, and its tempered ensemble versions ``swap'', ``random cr'', and ``augmented cr''.

Since it is the sampling efficiency of the latent chains $\boldX$ in the FHMM rather than associated parameters that is the direct target of our sampler, we fixed $\boldw$ value to $(0.075, 0.125, 0.15, 0.175, 0.2, 0.275)$ in these experiments. As a result, all samplers would be exploring the same conditional posterior, and we are able to directly compare the subclonal configurations identified by various sampling algorithms. Otherwise, joint updating of the weights $\boldw$ (though entirely feasible) would lead to label swapping effects and the possibility of samplers exploring entirely different regimes that then make direct comparisons across sampling methods more challenging. 

Figure \ref{fig:real_traces} shows the log-posterior traces and the traces of $\boldX$ for selected chromosomes, when using ensembles of the HB$(r=3)$ sampler. After a burn-in period of 10~000 iterations, the ``augmented cr'' ensemble has identified a probable configuration of $\boldX$ and it continues to explore parts of the state space which have higher posterior probability than those identified by other samplers. 

The augmented sampler is much better at capturing the uncertainty in underlying latent configurations (see Figure~\ref{fig:real_traces}(b)). For example, the third row corresponds to a subpopulation which has an extra copy of chromosome~21, but there is uncertainty whether it co-occurs with a whole extra copy of chromosome~22. Examining chromosome~17, the single-chain HB sampler and the ``random cr'' ensemble have identified a more fragmented latent configuration, whereas ``swap'' and ``augmented cr'' have combined these fragments into an alternative, more probable explanation. In biological terms, this is important since the more fragmented configuration would suggest a highly genomically unstable cancer genome related to a loss of genome integrity checkpoint mechanisms, whilst the alternative suggests a more moderate degree of instability.

\section{Conclusion}

We introduce an ensemble MCMC method to improve poorly mixing samplers for FHMMs. This is achieved by combining parallel tempering and a novel exchange move between pairs of chains achieved through an auxiliary variable augmentation. The former introduces a chain which explores the space freely and does not get stuck, whereas the latter provides an efficient procedure to exchange information between a tempered chain and our target.  The proposed method is a general purpose ensemble MCMC approach, but its most natural application case are sequential models. Specifically, we see this most useful for a broad class of models assuming Markov structure, where the augmented crossover move can be carried out at a cheap extra computational cost. A natural extension of this work is to integrate our ensemble technique into a sampling scheme for targeting latent variables $\boldX$ \textit{and} parameters $\theta$ in a joint model $\pi(\boldX, \theta)$. More exploration could also be carried out to explore optimal strategies for selecting or adapting the temperature ladder. However, our analyses suggest that for any given temperature ladder, the suggested augmented crossovers outperform non-augmented, classic approaches.

\section*{Acknowledgements} 

KM is supported by a UK Engineering and Physical Sciences Research Council Doctoral Studentship. CY is supported by a UK Medical Research Council Research Grant (Ref: MR/P02646X/1) and by The Alan Turing Institute under the EPSRC grant EP/N510129/1’.

\bibliography{main}

\begin{thebibliography}{23}
\providecommand{\natexlab}[1]{#1}
\providecommand{\url}[1]{\texttt{#1}}
\expandafter\ifx\csname urlstyle\endcsname\relax
  \providecommand{\doi}[1]{doi: #1}\else
  \providecommand{\doi}{doi: \begingroup \urlstyle{rm}\Url}\fi

\bibitem[Andrieu et~al.(2003)Andrieu, De~Freitas, Doucet, and
  Jordan]{andrieu2003introduction}
Andrieu, Christophe, De~Freitas, Nando, Doucet, Arnaud, and Jordan, Michael~I.
\newblock An introduction to {MCMC} for machine learning.
\newblock \emph{Machine learning}, 50\penalty0 (1-2):\penalty0 5--43, 2003.

\bibitem[Betancourt(2017)]{betancourt2017conceptual}
Betancourt, Michael.
\newblock A conceptual introduction to {Hamiltonian} {Monte Carlo}.
\newblock \emph{arXiv preprint arXiv:1701.02434}, 2017.

\bibitem[Cazier et~al.(2014)Cazier, Rao, McLean, Walker, Wright, Jaeger,
  Kartsonaki, Marsden, Yau, Camps, et~al.]{cazier2014whole}
Cazier, J-B, Rao, SR, McLean, CM, Walker, AK, Wright, BJ, Jaeger, EEM,
  Kartsonaki, C, Marsden, L, Yau, C, Camps, C, et~al.
\newblock Whole-genome sequencing of bladder cancers reveals somatic cdkn1a
  mutations and clinicopathological associations with mutation burden.
\newblock \emph{Nature communications}, 5:\penalty0 3756, 2014.

\bibitem[Crouse et~al.(1998)Crouse, Nowak, and Baraniuk]{crouse1998wavelet}
Crouse, Matthew~S, Nowak, Robert~D, and Baraniuk, Richard~G.
\newblock Wavelet-based statistical signal processing using hidden {Markov}
  models.
\newblock \emph{IEEE Transactions on signal processing}, 46\penalty0
  (4):\penalty0 886--902, 1998.

\bibitem[Earl \& Deem(2005)Earl and Deem]{earl2005parallel}
Earl, David~J and Deem, Michael~W.
\newblock Parallel tempering: Theory, applications, and new perspectives.
\newblock \emph{Physical Chemistry Chemical Physics}, 7\penalty0 (23):\penalty0
  3910--3916, 2005.

\bibitem[Frellsen et~al.(2016)Frellsen, Winther, Ghahramani, and
  Ferkinghoff-Borg]{frellsen2016bayesian}
Frellsen, Jes, Winther, Ole, Ghahramani, Zoubin, and Ferkinghoff-Borg, Jesper.
\newblock Bayesian generalised ensemble {Markov} chain {Monte Carlo}.
\newblock In \emph{Artificial Intelligence and Statistics}, pp.\  408--416,
  2016.

\bibitem[Gao et~al.(2016)Gao, Davis, McDonald, Sei, Shi, Wang, Tsai, Casasent,
  Waters, Zhang, et~al.]{gao2016punctuated}
Gao, Ruli, Davis, Alexander, McDonald, Thomas~O, Sei, Emi, Shi, Xiuqing, Wang,
  Yong, Tsai, Pei-Ching, Casasent, Anna, Waters, Jill, Zhang, Hong, et~al.
\newblock Punctuated copy number evolution and clonal stasis in triple-negative
  breast cancer.
\newblock \emph{Nature Genetics}, 2016.

\bibitem[Geyer(1991)]{geyercomputing}
Geyer, CJ.
\newblock \emph{Computing Science and Statistics Proceedings of the 23
  Symposium on the Interface; American Statistical Association: New York; p
  156}, 1991.

\bibitem[Ghahramani et~al.(1997)Ghahramani, Jordan, and
  Smyth]{ghahramani1997factorial}
Ghahramani, Zoubin, Jordan, Michael~I, and Smyth, Padhraic.
\newblock Factorial hidden {Markov} models.
\newblock \emph{Machine learning}, 29\penalty0 (2-3):\penalty0 245--273, 1997.

\bibitem[Gilks \& Roberts(1996)Gilks and Roberts]{gilks1996strategies}
Gilks, Walter~R and Roberts, Gareth~O.
\newblock Strategies for improving {MCMC}.
\newblock \emph{{Markov} chain {Monte Carlo} in practice}, 6:\penalty0 89--114,
  1996.

\bibitem[Ha et~al.(2014)Ha, Roth, Khattra, Ho, Yap, Prentice, Melnyk,
  McPherson, Bashashati, Laks, et~al.]{ha2014titan}
Ha, Gavin, Roth, Andrew, Khattra, Jaswinder, Ho, Julie, Yap, Damian, Prentice,
  Leah~M, Melnyk, Nataliya, McPherson, Andrew, Bashashati, Ali, Laks, Emma,
  et~al.
\newblock Titan: inference of copy number architectures in clonal cell
  populations from tumor whole-genome sequence data.
\newblock \emph{Genome research}, 24\penalty0 (11):\penalty0 1881--1893, 2014.

\bibitem[Holland(1992)]{holland1992adaptation}
Holland, John~H.
\newblock \emph{Adaptation in natural and artificial systems: an introductory
  analysis with applications to biology, control, and artificial intelligence}.
\newblock MIT press, 1992.

\bibitem[Jasra et~al.(2007)Jasra, Stephens, and Holmes]{jasra2007population}
Jasra, Ajay, Stephens, David~A, and Holmes, Christopher~C.
\newblock On population-based simulation for static inference.
\newblock \emph{Statistics and Computing}, 17\penalty0 (3):\penalty0 263--279,
  2007.

\bibitem[Kirkpatrick et~al.(1983)Kirkpatrick, Gelatt, Vecchi,
  et~al.]{kirkpatrick1983optimization}
Kirkpatrick, Scott, Gelatt, C~Daniel, Vecchi, Mario~P, et~al.
\newblock Optimization by simulated annealing.
\newblock \emph{Science}, 220\penalty0 (4598):\penalty0 671--680, 1983.

\bibitem[Liang \& Wong(2000)Liang and Wong]{liang2000evolutionary}
Liang, Faming and Wong, Wing~Hung.
\newblock Evolutionary {Monte Carlo}: Applications to cp model sampling and
  change point problem.
\newblock \emph{Statistica sinica}, pp.\  317--342, 2000.

\bibitem[Marchini \& Howie(2010)Marchini and Howie]{marchini2010genotype}
Marchini, Jonathan and Howie, Bryan.
\newblock Genotype imputation for genome-wide association studies.
\newblock \emph{Nature Reviews Genetics}, 11\penalty0 (7):\penalty0 499--511,
  2010.

\bibitem[Neal(2011)]{neal2011mcmc}
Neal, Radford~M.
\newblock {MCMC} using ensembles of states for problems with fast and slow
  variables such as gaussian process regression.
\newblock \emph{arXiv preprint arXiv:1101.0387}, 2011.

\bibitem[Rabiner \& Juang(1986)Rabiner and Juang]{rabiner1986introduction}
Rabiner, Lawrence and Juang, B.
\newblock An introduction to hidden {Markov} models.
\newblock \emph{{IEEE} {ASSP} Magazine}, 3\penalty0 (1):\penalty0 4--16, 1986.

\bibitem[Scott(2002)]{scott2002bayesian}
Scott, Steven~L.
\newblock Bayesian methods for hidden {Markov} models.
\newblock \emph{Journal of the American Statistical Association}, 2002.

\bibitem[Shestopaloff \& Neal(2014)Shestopaloff and
  Neal]{shestopaloff2014efficient}
Shestopaloff, Alexander~Y and Neal, Radford~M.
\newblock Efficient bayesian inference for stochastic volatility models with
  ensemble {MCMC} methods.
\newblock \emph{arXiv preprint arXiv:1412.3013}, 2014.

\bibitem[Titsias \& Yau(2014)Titsias and Yau]{titsias2014hamming}
Titsias, Michalis~K and Yau, Christopher.
\newblock Hamming ball auxiliary sampling for factorial hidden {Markov} models.
\newblock In \emph{Advances in Neural Information Processing Systems}, pp.\
  2960--2968, 2014.

\bibitem[Titsias \& Yau(2017)Titsias and Yau]{titsias2017hamming}
Titsias, Michalis~K and Yau, Christopher.
\newblock The hamming ball sampler.
\newblock \emph{Journal of the American Statistical Association}, pp.\  1--14,
  2017.

\bibitem[Yau(2013)]{yau2013oncosnp}
Yau, Christopher.
\newblock {OncoSNP-SEQ}: a statistical approach for the identification of
  somatic copy number alterations from next-generation sequencing of cancer
  genomes.
\newblock \emph{Bioinformatics}, 29\penalty0 (19):\penalty0 2482--2484, 2013.

\end{thebibliography}
\bibliographystyle{icml2018}






\newpage 

\begin{appendices}

\section*{Appendices}

\section{Pseudocode for the auxiliary variable crossovers}

\begin{algorithm} 
\caption{One-point crossover at point $t$} 
\begin{algorithmic} 
\Function{\textsc{Crossover}}{$(x_{1:T}, y_{1:T}, t)$}
\State $u_{1:T} \gets (y_1, \ldots, y_t, x_{t+1}, \ldots, x_T)$
\State $v_{1:T} \gets (x_1, \ldots, x_t, y_{t+1}, \ldots, y_T)$
\State \textbf{return}$(u_{1:T}, v_{1:T})$
\EndFunction
\end{algorithmic} \label{pseudocode:crossover}
\end{algorithm}

\begin{algorithm}
\caption{Scheme for an auxiliary variable two-point crossover between $\boldx_i$ and $\boldx_j$} 
\begin{algorithmic}
\State Pick $t$ uniformly $t \sim U(\{1, \ldots, T\})$ 
\State \# Flip a coin to decide the direction of crossover
\If{$u < 0.5$ where $u \sim U(0, 1)$}
\State $(\boldu, \boldv) \gets \textsc{crossover}(\boldx_i, \boldx_j, t)$
\Else 
\State $(\boldv, \boldu) \gets \textsc{crossover}(\boldx_i, \boldx_j, t)$
\EndIf
\State \# consider all normal and flipped crossovers of $u$ and $v$
\For{$t \in \{1, \ldots, T\}$} 
\State \# Normal crossover of $u$ and $v$
\State $(\boldz_i, \boldz_j) \gets \textsc{crossover}(\boldu, \boldv, t)$ 
\State $a_t \gets \pi_i(\boldz_i) \pi_j(\boldz_j)$
\State \# Flipped crossover of $u$ and $v$
\State $(\boldz_j, \boldz_i) \gets \textsc{crossover}(\boldu, \boldv, t)$ 
\State $a_{T+t} \gets \pi_i(\boldz_j) \pi_j(\boldz_i)$
\EndFor
\State \# Normalise the probabilities
\State $a_t \gets a_t / \sum_s a_s$ 
\State \# Pick index $t_0$ with probability proportional to $a_{t_0}$
\State $t_0 \sim \text{Discrete}(a_1, \ldots, a_{T}, a_{T+1}, \ldots, a_{2T})$ 
\If{$t_0 \le T$}
\State $(\boldx_i, \boldx_j) \gets \textsc{crossover}(\boldx_i, \boldx_j, t_0)$
\Else
\State $(\boldx_j, \boldx_i) \gets \textsc{crossover}(\boldx_i, \boldx_j, t_0)$
\EndIf
\end{algorithmic} \label{pseudocode:crossover2}
\end{algorithm}

\vfill

\section{Supplementary Figures for the toy example}

\begin{figure}[H]
\includegraphics[width=\columnwidth]{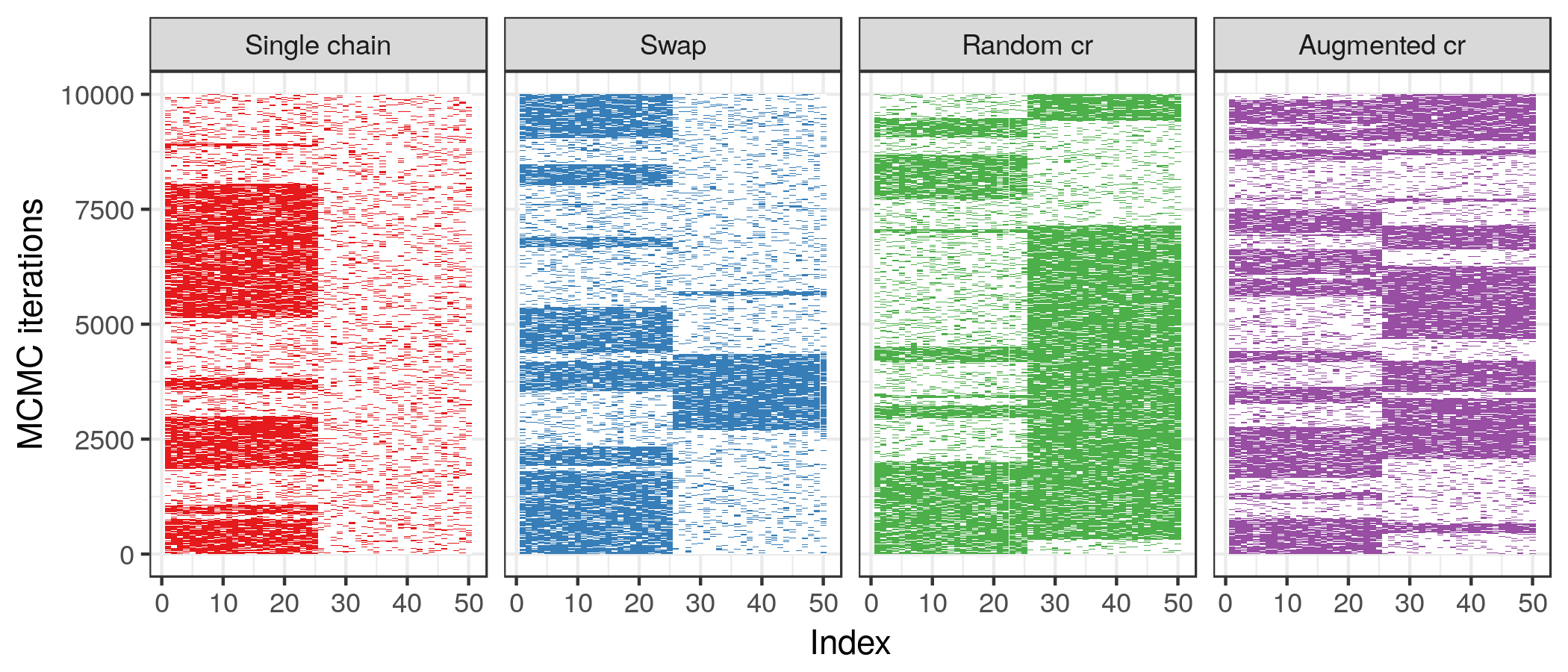}
\includegraphics[width=\columnwidth]{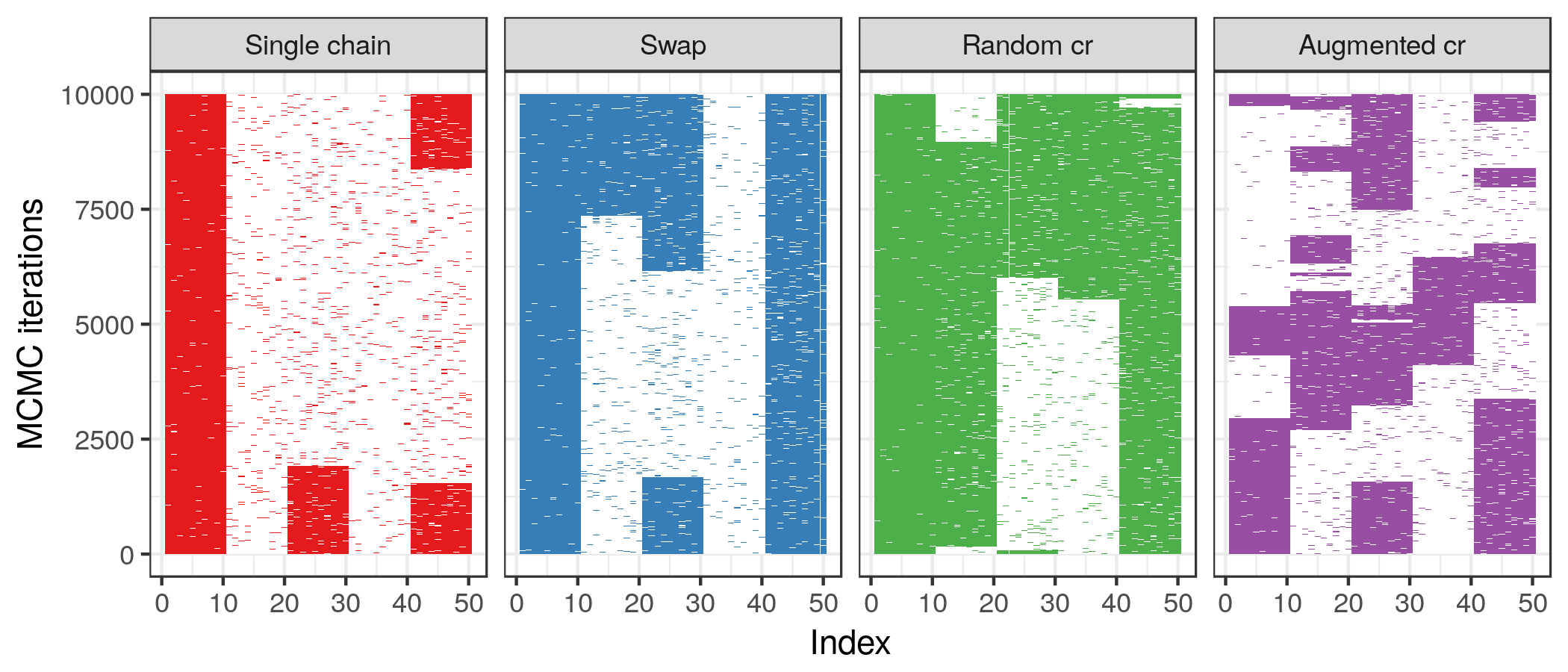}
\includegraphics[width=\columnwidth]{fig_toy/b10}
\caption{Heatmaps representing the trace plots of $\boldx$ for the experiment with $B \in \{2, 5, 10\}$ blocks, running a single chain Gibbs sampler (first panel), and its ensemble versions with various exchange moves: swap, random crossover, augmented crossover (in four panels). For each MCMC iteration, the elements of $\boldx$ have been colour coded: dark = 1, light = 0.}
\end{figure}


\end{appendices}

\end{document}